%% file: main.tex
\documentclass[11pt,letterpaper]{article}
\usepackage[margin=1in]{geometry}
\usepackage[T1]{fontenc}
\usepackage{lmodern,amsmath,amssymb,amsthm,mathtools,microtype}
\usepackage[colorlinks=true,linkcolor=blue,citecolor=blue,urlcolor=blue]{hyperref}
\newtheorem{theorem}{Theorem}[section]
\newtheorem{lemma}[theorem]{Lemma}
\newtheorem{proposition}[theorem]{Proposition}
\theoremstyle{definition}\newtheorem{definition}[theorem]{Definition}
\numberwithin{equation}{section}
\newcommand{\E}{\mathbb E}\newcommand{\PP}{\mathbb P}
\newcommand{\TV}{\operatorname{TV}}\newcommand{\Var}{\operatorname{Var}}
\hypersetup{pdftitle={A Fully Polynomial Randomized Approximation Scheme for Hypergraph Unreliability},pdfauthor={Zhijie Zheng}}
\title{A Fully Polynomial Randomized Approximation Scheme\\for Hypergraph Unreliability}
\author{Zhijie Zheng\\Peking University}\date{}
\begin{document}
\maketitle
\begin{abstract}
We give a fully polynomial randomized approximation scheme for all-terminal unreliability in explicitly represented undirected hypergraphs of arbitrary rank. The algorithm estimates the probability that independent hyperedge failures disconnect the hypergraph, allowing a different rational failure probability for every edge. For a common failure probability, this improves the quasi-polynomial approximation schemes of Cen, Li, and Panigrahi to polynomial time and answers the polynomial-time approximation question reiterated in their subsequent work on graphs.

The construction turns their representative-survival bound into an importance sampler whose relative second moment is bounded solely in terms of the number of vertices. We protect one randomly chosen representative pair during contraction, correct the path likelihood, and divide by the number of pairs surviving at the end. The resulting unbiased estimator has relative second moment at most $\binom n2$, even when unreliability is exponentially small.

A minimum-cut normalization makes this estimator implementable with polynomially bounded precision: the residual likelihood becomes a survival probability in an auxiliary absorbing process, which admits rational discretization with a fixed random-bit budget. Logarithmic rate rounding and an exact weighted minimum cut extend the construction to unequal rational failure probabilities. The algorithm's worst-case running time and random-bit usage are polynomial in the input length, the inverse relative error, and $\log(1/\delta)$ for confidence $1-\delta$.
\end{abstract}
\section{Introduction}
Network unreliability quantifies the risk that independent component failures disconnect a system. A hyperedge represents a component connecting an entire set of vertices, so one failure can remove a connection among arbitrarily many endpoints. We study \emph{all-terminal unreliability}, the probability that the surviving hypergraph has more than one connected component. A relative estimate measures this risk at its own scale, including when disconnection is exponentially unlikely.

For ordinary graphs, polynomial-time relative approximation has been available since Karger's FPRAS~\cite{Kar99}, and, for a common failure probability and fixed relative accuracy, the running time has recently reached almost linear in the number of edges~\cite{CLP25}. Extending these guarantees to unrestricted hyperedges encounters a structural obstacle. Near-minimum cuts can be enumerated in polynomial time at logarithmic rank for each fixed approximation factor, but their number can grow exponentially with the rank~\cite{KK15}. Enumerating these cuts can therefore require super-polynomial work at large rank.

Cen, Li, and Panigrahi~\cite{CLP24} gave quasi-polynomial approximation schemes for arbitrary-rank hypergraphs. Their later work~\cite{CLP25} again identified polynomial-time approximation in this model as an open problem. We resolve this question with an FPRAS. The result also allows independent rational failure probabilities that differ across edges, and bounds both running time and random-bit usage for every random outcome. The input length includes the number of hyperedges and all of their incidences.

\subsection{The result and the input model}
Let $H=(V,E)$ be an undirected hypergraph with explicitly listed vertices labeled by $[n]$. The edge collection $E$ is an explicit list of subsets of $V$; distinct labels may specify the same subset. Each labeled edge $e$ fails independently with rational probability $p_e$. Write $u_H(\mathbf p)$ for the probability of disconnection, and $u_H(p)$ when all failure probabilities equal $p$. Disconnection means that some nonempty proper vertex subset is crossed by no surviving edge. For $n\le1$ this event is empty. For $n\ge2$, connectivity means that the equivalence relation generated by membership in a common surviving edge has a single class.

Discard empty and singleton edges and remove duplicate incidences within an edge. Distinct parallel labels remain independent trials. After preprocessing, set
\[
m=|E|,\qquad I=\sum_{e\in E}|e|.
\]
The input length accounts for $n$, the incidences and labels, and the binary encodings of all rational parameters. Multiplicities are represented by their explicit edge lists.

\begin{theorem}\label{thm:main}
Given an explicitly represented undirected hypergraph $H$, independent rational edge-failure probabilities $p_e\in[0,1]$, and rational $\varepsilon,\delta\in(0,1)$, there is a randomized algorithm returning a rational $\widehat u\in[0,1]$ such that
\[
\PP\bigl[|\widehat u-u_H(\mathbf p)|\le\varepsilon u_H(\mathbf p)\bigr]\ge1-\delta.
\]
For every random outcome, its running time and random-bit usage are polynomial in the explicit input length, all probability encoding lengths, $\varepsilon^{-1}$, and $\log(1/\delta)$. The rank is unrestricted.
\end{theorem}

We first prove the common-probability case, which contains the central sampling construction. Section~\ref{sec:weighted} extends it to the full theorem.
\begin{theorem}\label{thm:common}
Given an explicitly represented undirected hypergraph $H$, a rational $p\in[0,1]$, and rational $\varepsilon,\delta\in(0,1)$, there is a randomized algorithm returning a rational $\widehat u\in[0,1]$ such that
\begin{equation}\PP\bigl[|\widehat u-u_H(p)|\le\varepsilon u_H(p)\bigr]\ge1-\delta.\label{eq:guarantee}\end{equation}
Its running time and number of random bits are bounded, for every random outcome, by a polynomial in the explicit input length, $\varepsilon^{-1}$, and $\log(1/\delta)$. The rank is unrestricted.
\end{theorem}

The theorem gives a polynomial-time relative estimate throughout the range of failure probabilities and hyperedge ranks. In particular, a large family of near-minimum cuts can coexist with efficient approximation of the total disconnection probability. The input model retains each hyperedge as one independent failure event, including parallel labels.

For the common-probability case, the algorithm estimates $u_H(p)/p^\lambda$, where $\lambda$ is the integer minimum cut. This normalized quantity lies between $1$ and $\binom n2$. Multiplication by the exactly represented rational $p^\lambda$ recovers a relative estimate of unreliability. This normalization allows a polynomial precision budget even when the probability being estimated is exponentially small.

\subsection{From representative survival to an algorithm}
The key is to sample disconnection through surviving representatives. Cen, Li, and Panigrahi's analysis bounds unreliability using only the original vertex pairs, even when the cut family is exponentially large. Turning that bound into an FPRAS requires a distribution under which a chosen pair survives, an exact correction for overlapping survival events, and a precision bound that remains polynomial for rare events.

\paragraph{Protect a pair and correct its multiplicity.}
Cen, Li, and Panigrahi~\cite[Section 3.2]{CLP24} represent independent failures by an exponential contraction process. Each block retains one original vertex as its representative. Two copies of each edge use distinct heads, giving a destruction rate of at least $\lambda$ for every surviving representative pair. We use this hazard bound to construct a proposal process: select one original pair uniformly and suppress every transition that would destroy it. All other transition rates are retained. A path under this process has an explicit likelihood relative to the original contraction process.

Every disconnected path under the original process has one or more surviving pairs. Dividing its contribution by that number corrects this multiplicity, so averaging over the initially selected pair counts each disconnected path exactly once. The hazard bound then gives the pointwise estimate $X\le\binom n2p^\lambda$. Together with $u_H(p)\ge p^\lambda$, this yields relative second moment at most $\binom n2$. The common head rule allows the multiplicity identity to be evaluated under a single original law.

\paragraph{Represent the likelihood by survival.}
Implementing the continuous-time sampler requires controlling numerical precision: both its horizon and its likelihood contain transcendental quantities. After dividing by $p^\lambda$, the remaining likelihood is the probability of avoiding an auxiliary absorbing state whose rate is $\kappa-\lambda\ge0$. The terminal reward is consequently either zero or the rational number $\binom n2/\binom K2$. A rational time grid and categorical draws with a fixed bit budget approximate the normalized mean to additive error at most $3\varepsilon/128$. Since that mean is at least one, standard batching gives a relative guarantee.

\paragraph{Allow different failure probabilities.}
For arbitrary $p_e$, the corresponding rates are $-\log p_e$. We round these rates to a common dyadic grid, prove multiplicative stability of unreliability under this rounding, and compute the resulting weighted minimum cut exactly. The same absorbing process runs to the rational horizon $1$. A rational relative approximation to the final exponential normalization completes the bit-model algorithm.

\subsection{Relation to previous work}
The exponential contraction representation, paired heads, and representative survival inequality are due to Cen, Li, and Panigrahi~\cite{CLP24}. Their Claim 3.6 supplies the hazard bound underlying $p^\lambda\le u_H(p)\le n^2p^\lambda$. We prove the needed form with original edges counted once and each copy running at rate $1/2$. Our contribution is to turn this analytical structure into a pair-preserving sampling process, establish its polynomial relative second moment, and implement it with bounded polynomial resources.

Importance sampling also underlies the subquadratic graph algorithm of Cen, He, Li, and Panigrahi~\cite{CHLP24}. Their estimator samples cuts through a spanning-tree packing, together with sparsification and recursive contraction. Here the sampling space consists of contraction paths that preserve a representative pair; the paired-head hazard controls their likelihoods for hyperedges of every rank.

The path change of measure is the finite-state Feynman--Kac identity. Reciprocal multiplicity correction is a standard importance-sampling technique, exemplified by the coverage estimator of Karp, Luby, and Madras~\cite[Section 5]{KLM89}. The paired-head hazard bound controls the estimator and ensures that the normalized absorption rate is nonnegative.

State-dependent importance sampling has also been used to estimate network unreliability. L'Ecuyer, Rubino, Saggadi, and Tuffin~\cite{LRST11} use minimum-cut approximations to guide sequential edge revelations and correct them by likelihood ratios. Their analysis allows nonuniform edge probabilities and establishes bounded relative error for a fixed topology in a rare-event limit. Our guarantee is uniform over explicitly represented arbitrary-rank hypergraphs and supplies a polynomial-size relative-variance bound together with worst-case polynomial bit complexity.

Guo and Jerrum~\cite{GJ19} use cluster popping to approximate all-terminal reliability, the probability that an ordinary graph remains connected. Feng, Fu, and Guo~\cite{FGG26} give an FPRAS for two-terminal connectivity in directed and undirected graphs and prove $\#\mathrm{BIS}$-hardness for undirected two-terminal unreliability. Our target is the probability that the surviving hypergraph has at least two connected components.

\section{Cuts and polynomial preprocessing}
For $\varnothing\ne S\subsetneq V$, let
\[
\partial_H S=\{e\in E:e\cap S\ne\varnothing,\ e\setminus S\ne\varnothing\}.
\]
Its cardinality counts edge labels, including parallel edges. If $n\ge2$, define $\lambda=\min_{\varnothing\ne S\subsetneq V}|\partial_H S|$. For connected $H$, $1\le\lambda\le m$. Failure of every edge in one minimum cut implies disconnection, so
\begin{equation}u_H(p)\ge p^\lambda\qquad(0<p<1).\label{eq:lower}\end{equation}
Lawler's directed cut representation~\cite{Law73} gives the following preprocessing lemma.
\begin{lemma}\label{lem:mincut}
The integer $\lambda$ is computable in polynomial time in the explicit hypergraph input length.
\end{lemma}
\begin{proof}
Fix an original vertex $r$. Form a directed graph with the original vertices and two auxiliary vertices $a_e,b_e$ per edge label. Insert $a_e\to b_e$ with capacity $1$, and, for each incidence $v\in e$, insert $v\to a_e$ and $b_e\to v$ with capacity $m+1$.

A vertex bipartition with source side $S$ extends to a directed cut of capacity $|\partial_H S|$: place both auxiliary vertices of an edge contained in one side on that side, and put $a_e$ on the source side and $b_e$ on the other side for a crossing edge. No capacity-$(m+1)$ arc crosses. Conversely, a directed cut of capacity at most $m$ crosses no such arc. A crossing hyperedge then forces $a_e$ onto the source side and $b_e$ onto the other side, so its unit arc crosses. Thus a minimum directed $r$--$v$ cut equals the minimum hypergraph cut separating these vertices.

Minimize over $v\ne r$. Every nontrivial cut separates $r$ from some such $v$, so this gives $\lambda$. The gadget has $n+2m$ vertices and $m+2I$ arcs. Integral augmenting-path maximum flow suffices: the flow value is at most $m$, every augmentation increases it by at least one, and each residual-path search takes polynomial time. At termination, the vertices reachable from the source in the residual graph define a cut of the same value. The $n-1$ computations therefore have polynomial bit complexity.
\end{proof}

The boundary cases can be resolved exactly. Return $0$ if $n\le1$. For $n\ge2$, return $1$ if the input is disconnected. For a connected input, return $0$ at $p=0$ and $1$ at $p=1$. Connectivity is tested by merging the vertices of each edge. Henceforth,
\begin{equation}n\ge2,\quad H\text{ is connected},\quad0<p<1,\quad1\le\lambda\le m.\label{eq:nondegenerate}\end{equation}

\section{A common representative contraction process}
A marked state $x$ is a partition of $V$ into blocks, each with one distinguished original vertex, its \emph{representative}. Initially the blocks are singletons. Write $R(x)$ for the representative set and $K(x)=|R(x)|$ for the number of blocks.

For each original edge $e$, maintain two labeled copies $(e,0)$ and $(e,1)$. Their current endpoints are the blocks meeting $e$. An edge with at most one endpoint is a loop in the quotient and remains a loop after all subsequent contractions. For a nonloop, copy $0$ has the endpoint with the smallest representative as head, and copy $1$ has the endpoint with the largest representative as head. A contraction merges all endpoints, retains the head representative, and deletes the others. The two heads are distinct and are recomputed after every contraction. This rule depends only on the marked state and copy label, so every pair-specific proposal below uses the same original process.

\begin{definition}
The original process $P$ gives each nonloop labeled copy contraction rate $1/2$. Rates add when distinct labels induce the same marked-state transition. Loops induce no state transition.
\end{definition}
Equivalently, the copies have independent rate-$1/2$ exponential first-arrival times. The first arrival for an edge contracts it, after which both copies are loops. Independent Poisson clocks give the same process: all rings after an edge becomes a loop have no effect. Thus every current nonloop copy retains its original rate after earlier contractions.

\begin{lemma}\label{lem:components}
At $T=\log(1/p)$, the blocks under $P$ have the distribution of the connected components after independent edge failures with probability $p$. In particular,
\begin{equation}\PP_P[K(X_T)\ge2]=u_H(p).\label{eq:components}\end{equation}
\end{lemma}
\begin{proof}
The minimum of the two independent rate-$1/2$ arrival times of an original edge has rate $1$, so the probability of no arrival by $T$ is $e^{-T}=p$, independently across edge labels. For any set of arrived edges, contraction in any order gives their connected-component partition. The retained representatives affect the marks but not this partition.
\end{proof}
Each effective contraction decreases $K$, so there are at most $n-1$ effective jumps. Deleted representative labels never return; survival of an original pair is therefore a monotone path event.

\section{The paired hazard bound and safe suppression}
Fix $w=\{a,b\}\in\binom V2$. While both labels survive, their blocks are distinct. A copy contraction is \emph{safe} if it retains both labels, and \emph{killing} otherwise. If $c_w(x)$ and $d_w(x)$ are the respective numbers of nonloop copies, write
\begin{equation}s_w(x)=c_w(x)/2,\qquad\kappa_w(x)=d_w(x)/2.\label{eq:rates}\end{equation}
If $\ell(x)$ counts original nonloop edge labels, then
\begin{equation}s_w(x)+\kappa_w(x)=\ell(x)\le m.\label{eq:totalrate}\end{equation}
\begin{lemma}[Paired representative hazard]\label{lem:hazard}
In every marked quotient state in which $a$ and $b$ both survive,
\begin{equation}\kappa_w(x)\ge\lambda.\label{eq:hazard}\end{equation}
\end{lemma}
\begin{proof}
Let $A,B$ be the blocks represented by $a,b$, and let $D_A,D_B$ count original edge labels crossing their block cuts. Both blocks are nonempty proper subsets, so $D_A,D_B\ge\lambda$. An edge incident to exactly one protected block has that block as a non-head endpoint in at least one copy, contributing at least $1/2$ to the killing rate. An edge incident to both blocks destroys at least one protected representative under either head choice, contributing $1$. Summing gives
\[\kappa_w(x)\ge\tfrac12(D_A+D_B)\ge\lambda.\]
The factor $1/2$ is the copy rate. This is the paired-head argument of~\cite[Claim 3.6]{CLP24} with original edge labels counted once.
\end{proof}
\begin{definition}\label{def:safe}
The safe process $Q_w$ retains each safe copy contraction at rate $1/2$ and suppresses each killing copy contraction. Suppressed copies remain available: at every new state, all original labels are reexamined and each currently safe nonloop copy again has rate $1/2$.
\end{definition}
Under $Q_w$ the protected pair always survives, so $K\ge2$ and there are at most $n-2$ effective contractions. Its holding rate is $s_w(x)$. When this rate is zero the state remains fixed, and the likelihood continues to accumulate through the terminal horizon. This includes a path that has already reached two blocks. These conventions specify the proposal law completely and will also be used in its finite-bit implementation.

\section{Likelihood and inverse multiplicity}
The two ingredients of the estimator are a likelihood correction for protecting a pair and a multiplicity correction for the pairs surviving on one path. We derive them separately.
\begin{lemma}[Safe path identity]\label{lem:path}
For every bounded measurable path functional $F$ on paths retaining $w$ through time $T$,
\begin{equation}
\E_{Q_w}\!\left[e^{-\int_0^T\kappa_w(X_t)\,dt}F\right]
=\E_P\!\left[\mathbf1_{\{w\subseteq R(X_T)\}}F\right].\label{eq:path}
\end{equation}
The right-hand product is zero when $w$ does not survive.
\end{lemma}
\begin{proof}
Fix a safe labeled sequence with states $x_0,\ldots,x_j$ and jump times $0<t_1<\cdots<t_j<T$. Set $t_0=0$, $t_{j+1}=T$, and $\Delta_i=t_{i+1}-t_i$. The two laws assign the same rate to each listed jump. Their densities differ only in the probability of waiting between these jumps and from the last jump to the observation time. The original density for these jumps and no further effective jump before $T$ is
\[2^{-j}\exp\!\left(-\sum_{i=0}^j[s_w(x_i)+\kappa_w(x_i)]\Delta_i\right),\]
whereas the safe density is
\[2^{-j}\exp\!\left(-\sum_{i=0}^js_w(x_i)\Delta_i\right).\]
Their ratio is the weight in~\eqref{eq:path}. The term $i=j$ accounts for the final holding interval, including when $j=0$ or the safe exit rate vanishes. For example, if $s_w(x_j)=0$, the proposal stays at $x_j$ until $T$, while the original process avoids a killing jump during that interval with probability $e^{-\kappa_w(x_j)(T-t_j)}$. This factor remains part of the ratio even though the proposal makes no further jump.

Summing over safe sequences and integrating over their jump times gives the identity. There are at most $n-2$ jumps. The labels distinguish transitions that produce the same state change. Summing their rates gives the same identity for unlabeled paths. Since deleted representatives never return, these are exactly the original paths retaining $w$ at time $T$.
\end{proof}

Set $N=\binom n2$. Choose $w$ uniformly from the $N$ initial pairs, sample $Q_w$ to time $T$, and define
\begin{equation}X=\frac{N}{\binom{K(X_T)}2}\exp\!\left(-\int_0^T\kappa_w(X_t)\,dt\right).\label{eq:estimator}\end{equation}
Its denominator is positive on every proposal path.
\begin{theorem}\label{thm:estimator}
The estimator~\eqref{eq:estimator} satisfies
\begin{equation}\E X=u_H(p),\qquad0\le X\le Np^\lambda,\qquad\frac{\E X^2}{u_H(p)^2}\le N.\label{eq:moment}\end{equation}
\end{theorem}
\begin{proof}
Apply Lemma~\ref{lem:path} with $F=1/\binom{K(X_T)}2$ and average over the initial pair:
\begin{equation}
\E X=\sum_{w\in\binom V2}\E_P\!\left[\frac{\mathbf1_{\{w\subseteq R(X_T)\}}}{\binom{K(X_T)}2}\right]
=\PP_P[K(X_T)\ge2]=u_H(p).\label{eq:multiplicity}
\end{equation}
On a path ending with $K\ge2$ representatives, exactly $\binom K2$ initial pairs survive. On a path ending with one block, every summand is zero. Lemma~\ref{lem:path} has expressed every summand under the same law $P$, so the sum can be evaluated on one marked trajectory. The final representative set determines both the surviving pairs and their common denominator. Because the head rule is common to all pairs, this count applies to every choice of the initially protected pair.

Lemma~\ref{lem:hazard} gives integrated hazard at least $\lambda T$, while the denominator in~\eqref{eq:estimator} is at least one. Hence $X\le Ne^{-\lambda T}=Np^\lambda$, and $X^2\le Np^\lambda X$. Taking expectations and using~\eqref{eq:lower} proves the last bound.
\end{proof}
The relative variance of the ideal estimator is at most $N-1$. The proof also gives $u_H(p)\le Np^\lambda$ directly from representative pairs. The next two sections convert this estimator into a procedure with a bounded number of rational operations and random bits.

\section{Normalization by an absorbing process}
Write
\begin{equation}\mu=\frac{u_H(p)}{p^\lambda},\qquad1\le\mu\le N.\label{eq:mu}\end{equation}
The normalized version of~\eqref{eq:estimator} is
\[Z=\frac{N}{\binom{K(X_T)}2}\exp\!\left(-\int_0^T[\kappa_w(X_t)-\lambda]\,dt\right).\]
The residual rate in the exponent is nonnegative, so the corresponding weight can be realized as a survival probability.
\begin{definition}
For each protected pair $w$, the process $R_w$ retains every safe contraction at rate $1/2$ and has a transition from each live state $x$ to an absorbing cemetery state $\dagger$ at rate
\begin{equation}f_w(x)=\kappa_w(x)-\lambda.\label{eq:cemetery}\end{equation}
Its terminal reward is
\begin{equation}g(x)=N/\binom{K(x)}2\quad\text{for live }x,\qquad g(\dagger)=0.\label{eq:reward}\end{equation}
\end{definition}
The cemetery rate is nonnegative by Lemma~\ref{lem:hazard}, and the live exit rate is
\begin{equation}s_w(x)+f_w(x)=\ell(x)-\lambda\le m.\label{eq:exit}\end{equation}
All rates are half-integers.
\begin{proposition}\label{prop:normalized}
Choose $w$ uniformly and run $R_w$ to time $T$. Then $Y=g(X_T)$ satisfies
\begin{equation}\E Y=\mu,\qquad0\le Y\le N,\qquad\E Y^2\le N\mu.\label{eq:normalized}\end{equation}
\end{proposition}
\begin{proof}
For a safe path, its density under $R_w$ without absorption equals its $Q_w$ density times $\exp(-\int_0^T f_w(X_t)\,dt)$. The holding-density calculation in Lemma~\ref{lem:path}, including the terminal interval, proves this identity. Equivalently, one can first sample a $Q_w$ path and then add an absorption clock with the piecewise constant rate $f_w$ along it. Conditional on that path, the probability of reaching $T$ without absorption is exactly the displayed exponential. Multiplying this probability by the live terminal reward gives $Z$, the normalized likelihood estimator. Absorbed paths have reward zero. Averaging over $w$ therefore gives $\E Y=\E Z=\E X/p^\lambda=\mu$. Finally $Y^2\le NY$.
\end{proof}
The variance bound $\Var(Y)\le N\mu$ and the lower bound $\mu\ge1$ allow a small additive bias from finite precision while preserving relative accuracy.

\input{finite_bits}
\input{weighted}
\input{witnesses}

\end{document}

%% file: finite_bits.tex
\section{A bounded finite-bit sampler}
Write $p=A/B$ with integers $0<A<B$, let $b$ be the binary length of $B$, and put $H_0=b+1$. The fraction need not be reduced. Since $A\ge1$,
\begin{equation}T=\log(B/A)\le\log B<b<H_0.\label{eq:horizon}\end{equation}
Logarithms are natural unless a base is shown. We control three errors: approximating the horizon, discretizing time, and sampling rational categorical probabilities with a fixed number of bits.

\subsection{A rational approximation to the horizon}
\begin{lemma}\label{lem:log}
For every positive rational $\eta$, a nonnegative rational $T'$ with
\[
0\le T'\le T<H_0,\qquad T-T'\le\eta
\]
can be computed in time polynomial in the encodings of $A,B,\eta$ and in $\log(1/\eta)$ when $\eta<1$.
\end{lemma}
\begin{proof}
Integer comparisons find $j\ge0$ with $2^j\le B/A<2^{j+1}$. Then $j\le b$ and $y=B/(2^jA)\in[1,2)$. For $x\in[1,2]$, let $z=(x-1)/(x+1)\in[0,1/3]$ and
\[
L_J(x)=2\sum_{i=0}^{J-1}\frac{z^{2i+1}}{2i+1}.
\]
Integrating the geometric series gives
\[
\log x=2\sum_{i=0}^\infty\frac{z^{2i+1}}{2i+1}.
\]
The nonnegative tail is at most $2z^{2J+1}/(1-z^2)\le3\cdot9^{-J}$. Choose the smallest integer $J\ge1$ with $3(j+1)9^{-J}\le\eta$, and set $T'=jL_J(2)+L_J(y)$. This lower approximation to $T=j\log2+\log y$ has error at most $\eta$.

There are $O(1+\log(b+1)+\max\{0,\log(1/\eta)\})$ terms. The rational bases have polynomial bit length, and exact rational summation has polynomial bit complexity. The product of all term denominators has polynomially many bits in $b$ and $J$.
\end{proof}
For the algorithm, choose
\begin{equation}
\eta=\frac{\varepsilon}{128Nm},\qquad
L=\left\lceil\frac{128Nm^2H_0^2}{\varepsilon}\right\rceil,\qquad h=\frac{T'}L.
\label{eq:grid}
\end{equation}
Then $mh\le1$. The horizon tolerance makes $m\eta\le\varepsilon/(128N)$, and the choice of the number of time steps ensures $m^2H_0^2/L\le\varepsilon/(128N)$. These are the two total variation budgets proved below. The factor $N$ accounts for the largest possible terminal reward when converting a distributional error into an expectation error. All precision parameters are found by exact rational comparisons.

\subsection{Rational Euler transitions}
For a live state $x$ and a fixed protected pair, one Euler step assigns probabilities
\begin{equation}
\begin{aligned}
h/2& &&\text{to each safe labeled copy contraction},\\
h[\kappa_w(x)-\lambda]& &&\text{to the cemetery},\\
1-h[s_w(x)+\kappa_w(x)-\lambda]& &&\text{to staying at }x.
\end{aligned}\label{eq:euler}
\end{equation}
These are nonnegative rationals summing to one by~\eqref{eq:exit}. The cemetery remains absorbing. There are at most $2m+2$ categories; categories of probability zero may be omitted. A self-loop preserves the availability of every copy. After a contraction, recompute the quotient endpoints, heads, and safe/killing classification. The current partition, representatives, and incidence list suffice for each update, irrespective of the size of the full state space.

\begin{lemma}\label{lem:euler}
With exact categorical draws, the total variation distance between $L$ Euler steps and $R_w$ at time $T'$ is at most
\begin{equation}\frac{m^2(T')^2}{L}\le\frac{\varepsilon}{128N}.\label{eq:eulererror}\end{equation}
The distance between $R_w$ at times $T'$ and $T$ is at most $m|T-T'|\le\varepsilon/(128N)$.
\end{lemma}
\begin{proof}
Uniformize $R_w$ at rate $m$. At each ring of the Poisson clock, apply a state-dependent kernel $P_w$: each safe copy has probability $1/(2m)$, the cemetery has probability $f_w(x)/m$, and the residual mass is a self-loop. Equation~\eqref{eq:exit} makes this a stochastic kernel; it is the identity at the cemetery. Thinning recovers the required rates. Thus the exact transition over time $h$ is a mixture of $P_w^q$ with $q\sim\operatorname{Poisson}(mh)$, and the Euler transition uses $q\sim\operatorname{Bernoulli}(mh)$.

For $0\le a\le1$, comparing the masses at $0,1$, and at least $2$ gives
\[
\TV(\operatorname{Poisson}(a),\operatorname{Bernoulli}(a))
=a(1-e^{-a})\le a^2.
\]
Applying a common kernel contracts total variation. The one-step discrepancy is therefore at most $(mh)^2$, uniformly in the state. To accumulate the errors, couple the two chains whenever their current states agree. Conditional on agreement up to a given step, the probability of their first disagreement at that step is at most $(mh)^2$. Summing over the $L$ possible first-disagreement steps bounds the total discrepancy by $L(mh)^2=m^2(T')^2/L$, which satisfies~\eqref{eq:eulererror} by~\eqref{eq:grid}. The uniform bound makes this coupling valid even though transition probabilities depend on the current state.

To compare the two horizons, couple the processes along a common uniformized trajectory up to the shorter horizon. The states agree unless a ring occurs in the remaining interval, whose probability is at most $1-e^{-m|T-T'|}\le m|T-T'|$.
\end{proof}

\subsection{Categorical draws with a fixed bit budget}
Choose the smallest integer $q\ge1$ such that
\begin{equation}2^q\ge\frac{128N[N+L(2m+2)]}{\varepsilon}.\label{eq:bits}\end{equation}
This choice can be made by repeated doubling and exact rational comparison. For a rational categorical distribution, use $q$ fair bits to generate a uniform integer $J\in\{0,\ldots,2^q-1\}$ and select the half-open cumulative interval containing $J/2^q$. Comparisons with rational endpoints are performed exactly by cross multiplication. Use the same procedure for the initially uniform distribution on $N$ pairs.

\begin{lemma}\label{lem:categorical}
For a distribution with at most $C$ categories, this draw has total variation error at most $C2^{-q}$. The error over the initial pair and at most $L$ updates is at most $\varepsilon/(128N)$.
\end{lemma}
\begin{proof}
Couple $U\sim\operatorname{Unif}[0,1)$ with $\lfloor2^qU\rfloor/2^q$. The selected categories agree unless a cumulative boundary lies between them. For each boundary, the set of draws that cross it has measure at most $2^{-q}$, which gives the bound $C2^{-q}$. A boundary exactly on the grid changes only a null set in the continuous draw.

Initially there are $N$ categories. At each later step, condition on the past and couple the draws when their states agree. There are at most $2m+2$ categories, so a union bound on the step of the first discrepancy gives $[N+L(2m+2)]2^{-q}\le\varepsilon/(128N)$. Absorption permits skipping the remaining updates.
\end{proof}

\subsection{The single-sample procedure}
Generate one normalized sample $\widehat Y$ as follows.
\begin{enumerate}
\item Use a fresh $q$-bit draw to select a pair $w$ from the uniform target distribution on $\binom V2$. Initialize the singleton state.
\item Perform $L$ Euler steps using fresh $q$-bit draws with probabilities~\eqref{eq:euler}. Each step contracts a safe copy, leaves the state unchanged, or enters the cemetery. Upon absorption, return $0$.
\item After $L$ live steps, return the rational reward $N/\binom K2$.
\end{enumerate}
The preprocessing values $\lambda,T',L,h,q$ are shared across independent samples. Both the number of steps and the number of random bits have fixed upper bounds.
\begin{proposition}\label{prop:finite}
For $\nu=\E\widehat Y$,
\begin{equation}
0\le\widehat Y\le N,\qquad
|\nu-\mu|\le\frac{3\varepsilon}{128},\qquad
\Var(\widehat Y)\le N\nu\le2N\mu.\label{eq:finite}
\end{equation}
\end{proposition}
\begin{proof}
Compare the exact process at $T$ with the exact process at $T'$, then with the exact Euler simulation, and finally with its fixed-bit implementation. Lemmas~\ref{lem:euler} and~\ref{lem:categorical} bound each total variation discrepancy by $\varepsilon/(128N)$. For a reward in $[0,N]$, the difference in expectations is at most $N$ times the total variation distance, proving the bias bound. The unchanged reward range gives $\E\widehat Y^2\le N\nu$. Since $\mu\ge1$ and $\varepsilon<1$, $\nu\le\mu+3\varepsilon/128\le2\mu$.
\end{proof}
The bias is additive in a normalized quantity with mean at least one. Thus the precision budget depends on $\varepsilon$ and the input parameters even when $p^\lambda$ is very small.

\section{Confidence amplification and bit complexity}
Set
\begin{equation}S=\left\lceil128N/\varepsilon^2\right\rceil.\label{eq:batches}\end{equation}
Let $r$ be the smallest positive odd integer with $2^{-r}\le\delta$. Repeated doubling finds $r=O(\log(2/\delta))$. Generate $r$ independent batches of $S$ samples, let $\overline Y_i$ be their exact rational averages, and let $M_*$ be their median. Return
\begin{equation}\widehat u=\min\{1,p^\lambda M_*\}.\label{eq:output}\end{equation}
All samples and categorical draws use independent random bits.
\begin{lemma}\label{lem:confidence}
The output~\eqref{eq:output} satisfies~\eqref{eq:guarantee}.
\end{lemma}
\begin{proof}
Proposition~\ref{prop:finite} and Chebyshev's inequality imply
\[
\PP\!\left[|\overline Y_i-\nu|>\frac{\varepsilon\mu}{2}\right]
\le\frac{4\Var(\widehat Y)}{S\varepsilon^2\mu^2}
\le\frac{8N}{S\varepsilon^2\mu}\le\frac1{16}.
\]
On the complementary event,
\[
|\overline Y_i-\mu|
\le\frac{\varepsilon\mu}2+\frac{3\varepsilon}{128}
\le\left(\frac12+\frac3{128}\right)\varepsilon\mu<\varepsilon\mu.
\]
If a strict majority of the batch averages lie in $[(1-\varepsilon)\mu,(1+\varepsilon)\mu]$, then the median also lies in that interval. Independence and a union bound over subsets of inaccurate batches bound the probability that a majority are inaccurate by $2^r(1/16)^{r/2}=2^{-r}\le\delta$. Multiplication by the positive exact rational $p^\lambda$ gives the desired relative guarantee. Clipping at $1$ cannot increase the distance to any probability in $[0,1]$.
\end{proof}

\begin{lemma}\label{lem:complexity}
The algorithm has bounded polynomial running time and random-bit usage in the parameters of Theorem~\ref{thm:common}.
\end{lemma}
\begin{proof}
Lemma~\ref{lem:mincut} handles preprocessing. Store a marked state as an array mapping each original vertex to its representative. Scanning the incidences of an edge identifies its quotient endpoints, smallest and largest representatives, and the representatives killed by each head choice. Boolean marking arrays or dictionaries suffice. A contraction updates the array in $O(n)$ operations. Thus one full scan and update uses $O(n+I)$ elementary combinatorial operations, apart from polynomial-bit arithmetic and indexing costs.

Lemma~\ref{lem:log} gives $T'$ with polynomial encoding length. The parameter bounds are
\[
\begin{aligned}
L&=O\!\left(\frac{Nm^2(b+1)^2}{\varepsilon}\right),&
S&=O(N\varepsilon^{-2}),\\
r&=O(\log(2/\delta)),&
q&=O\!\left(1+\log\frac{N[N+L(2m+2)]}{\varepsilon}\right).
\end{aligned}
\]
Each sample uses at most $L$ updates and $(L+1)q$ random bits. Transition probabilities are formed from $h$, half-integer rates bounded by $m$, and $\lambda$, so their bit lengths and the cost of exact cumulative comparisons are polynomial.

Each terminal reward has numerator and denominator at most $N$. Exact summation of $S$ rewards has denominator length at most $O(S\log(N+1))$, even if performed by multiplying all denominators. Averaging and median comparisons are therefore polynomial. Finally $p^\lambda=A^\lambda/B^\lambda$ has encoding length $O(mb)$ because $\lambda\le m$. Integer exponentiation and the final multiplication also have polynomial bit complexity.
\end{proof}
The total number of Euler updates is conservatively bounded by
\begin{equation}
O\!\left(N^2m^2(b+1)^2\varepsilon^{-3}\log(2/\delta)\right).
\label{eq:updatecount}
\end{equation}
Combining this count with the per-update work, rational arithmetic, and preprocessing gives a polynomial bound on the bit complexity. The discretization provides a direct implementation; faster implementations are discussed in Section~\ref{sec:discussion}.

\begin{proof}[Proof of Theorem~\ref{thm:common}]
Preprocessing gives the exact boundary answers. In the remaining cases, Lemma~\ref{lem:hazard} and~\eqref{eq:exit} ensure valid rates and probabilities. Lemma~\ref{lem:confidence} proves the approximation guarantee, and Lemma~\ref{lem:complexity} gives the worst-case bounds on running time and random-bit usage.
\end{proof}

%% file: weighted.tex
\section{Independent rational failure probabilities}\label{sec:weighted}
The same representative process handles unequal failure probabilities after replacing unit rates by logarithmic rates. The implementation requires two additional steps: relative stability under rate rounding, and exact weighted minimum-cut computation. These preserve relative accuracy even for rare disconnection events.

\subsection{Deterministic edges and relative stability}
Discard empty and singleton edges, and remove repeated incidences within each edge. Contract the components generated by edges with $p_e=0$, which are always present, and delete edges with $p_e=1$, which are always absent. Replace the remaining incidences by quotient endpoints, remove loops, and retain each parallel edge label as an independent trial. This preserves connectivity in every realization. Return $0$ if at most one quotient vertex remains, and $1$ if the remaining support is disconnected. Otherwise relabel the quotient by $[n]$. All subsequent parameters refer to this connected quotient, with $n\ge2$, $m\ge1$, and $0<p_e<1$. Write $u(\mathbf p)$ for its unreliability.

\begin{lemma}[Product-measure relative stability]\label{lem:stability}
Let $\mathcal A$ be an increasing event in independent failed-edge indicators. If $p_e,q_e\in[0,1]$ and
\[
e^{-t_e}p_e\le q_e\le e^{t_e}p_e\qquad(t_e\ge0),
\]
then
\[
e^{-\sum_e t_e}\PP_{\mathbf p}(\mathcal A)
\le\PP_{\mathbf q}(\mathcal A)
\le e^{\sum_e t_e}\PP_{\mathbf p}(\mathcal A).
\]
\end{lemma}
\begin{proof}
With all other coordinate distributions fixed, the event probability is $ap+b$ as a function of one failure probability, where
\[
a=\PP(\mathcal A\mid\text{edge failed})-\PP(\mathcal A\mid\text{edge present})\ge0,
\qquad b=\PP(\mathcal A\mid\text{edge present})\ge0.
\]
Here conditioning means fixing the state of this edge and sampling all others independently. This interpretation also defines both quantities when the failure probability is an endpoint. If $e^{-t}p\le q\le e^tp$, positivity gives
$e^{-t}(ap+b)\le aq+b\le e^t(ap+b)$. Change one coordinate at a time and multiply the inequalities.
\end{proof}

Set
\begin{equation}\tau=\frac{\varepsilon}{64m},\qquad r_e=-\log p_e.\label{eq:tau}\end{equation}
For $p_e=A_e/B_e$, let $b_e$ be the binary length of $B_e$. Lemma~\ref{lem:log} gives a rational lower bound $l_e\le r_e$ with $r_e-l_e\le\tau/4$. Put $u_e=l_e+\tau/4$. Choose the smallest nonnegative integer $a$ with $2^{-a}\le\tau/4$, and set
\begin{equation}d=2^{-a},\qquad r'_e=d\lceil u_e/d\rceil.\label{eq:roundedrates}\end{equation}
The first approximation gives $r_e\le u_e\le r_e+\tau/4$. Rounding $u_e$ upward to the dyadic grid adds less than $d\le\tau/4$, so the two errors together are at most $\tau/2$. Consequently,
\begin{equation}
0<r_e\le r'_e\le r_e+\tau/2,\qquad
R:=\sum_e r'_e\le\sum_e b_e+m\tau.\label{eq:ratesum}
\end{equation}
The rates have a common denominator and polynomial bit length; their total $R$ is also polynomially bounded in the input length. Upward rounding preserves positivity even for very small original rates.

Let $p'_e=e^{-r'_e}$ and $u'=u(\mathbf p')$ define a comparison product measure. The algorithm operates directly on the rational rates. Since disconnection is increasing in failures, Lemma~\ref{lem:stability} implies
\begin{equation}e^{-\varepsilon/64}u(\mathbf p)\le u'\le e^{\varepsilon/64}u(\mathbf p).\label{eq:stability}\end{equation}
Upward rounding also gives $u'\le u(\mathbf p)$; the symmetric bound above is convenient when combining errors. Indeed, each ratio $p'_e/p_e=e^{-(r'_e-r_e)}$ lies between $e^{-\tau/2}$ and $1$. The stability lemma accumulates these coordinatewise logarithmic errors by their sum, which is at most $m\tau/2=\varepsilon/128$. Equation~\eqref{eq:stability} uses the slightly larger budget $\varepsilon/64$. Thus the required precision is set by the number of edges and the requested relative error, including when the original unreliability is very small.

\subsection{An exact rational weighted minimum cut}
Define
\begin{equation}\Lambda=\min_{\varnothing\ne S\subsetneq V}\sum_{e\in\partial_H S}r'_e.\label{eq:weightedcut}\end{equation}
The connected support and positive rates give $0<\Lambda\le R$.
\begin{lemma}\label{lem:weightedcut}
The value $\Lambda$ can be computed exactly in polynomial time in the explicit input and rate encodings.
\end{lemma}
\begin{proof}
Use the gadget of Lemma~\ref{lem:mincut}, with capacity $r'_e$ on $a_e\to b_e$ and $R+1$ on incidence arcs. Every vertex cut extends to a cut in the gadget with the same weighted hyperedge-cut capacity. A minimum cut from the root to another original vertex has value at most $R$, so it crosses no incidence arc. The same argument then identifies its value with the corresponding weighted hypergraph cut. Minimize over the $n-1$ targets.

Scale by $2^a$ to obtain integer capacities of polynomial bit length. Use shortest-augmenting-path maximum flow. For the capacity-independent complexity bound, observe that breadth-first residual distances from the source never decrease after an augmentation. If a residual arc is saturated on a shortest path and is later saturated again in the same direction, an intervening path must have used its reverse, increasing the tail distance by at least two. Each directed residual arc can consequently be saturated on a shortest augmenting path only $O(v_*)$ times, where $v_*$ is the number of gadget vertices. Charging each augmentation to a saturated arc gives $O(v_*e_*)$ augmentations and $O(v_*e_*^2)$ elementary work, where $e_*$ counts gadget arcs. Arithmetic on the scaled capacities has polynomial bit complexity. Divide the final value by $2^a$.
\end{proof}
This capacity-independent augmentation bound keeps the running time polynomial in the bit length of the scaled capacities.

\subsection{The weighted normalized process}
Give the two copies of edge $e$ rate $r'_e/2$ and run the original process to horizon $1$, with the same minimum- and maximum-representative head rule. The first arrival has rate $r'_e$, independently across labels. Thus the component partition at time $1$ has the distribution induced by the failure probabilities $\mathbf p'$.

For a protected pair $w$, let $s_w(x),\kappa_w(x)$ be the total safe and killing rates. In the paired-head argument, an edge incident to exactly one protected block contributes at least $r'_e/2$ to the killing rate, and an edge incident to both blocks contributes $r'_e$. Hence, for protected blocks $A,B$,
\begin{equation}
\kappa_w(x)\ge\frac12\left(\sum_{e\in\partial_H A}r'_e+
\sum_{e\in\partial_H B}r'_e\right)\ge\Lambda.\label{eq:weightedhazard}
\end{equation}
The two block cuts are original nontrivial vertex cuts. Labels and rates remain attached to their original edges throughout contraction.

Suppress the killing transitions while keeping every safe rate $r'_e/2$ and leaving suppressed copies available. The density calculation of Lemma~\ref{lem:path} still applies: actual transition-rate factors agree, and the holding-time density factors have ratio $\exp(-\int_0^1\kappa_w(X_t)\,dt)$, including the final interval. Uniformly choosing the initial pair and dividing by $\binom K2$ therefore gives an unbiased estimator of $u'$.

For normalization, add a cemetery transition of rate $\kappa_w(x)-\Lambda$ while retaining the safe rates. The total live exit rate is at most $R$. Let $F=0$ at the cemetery and $F=N/\binom K2$ in a live terminal state. Proposition~\ref{prop:normalized} now gives
\begin{equation}
\E F=\mu'=e^\Lambda u',\qquad0\le F\le N,\qquad1\le\mu'\le N.
\label{eq:weightedmean}
\end{equation}
The lower bound follows by failing every edge of one weighted minimum cut, an event of probability $e^{-\Lambda}$. Every rate of the normalized process is rational and explicitly computable.

\subsection{Finite sampling and concentration}
Put
\begin{equation}
M_0=\max\{1,R\},\qquad
L'=\left\lceil\frac{1024NM_0^2}{\varepsilon}\right\rceil,\qquad h'=1/L'.
\label{eq:weightedgrid}
\end{equation}
An Euler step assigns probability $h'r'_e/2$ to a safe copy of $e$, probability $h'[\kappa_w(x)-\Lambda]$ to the cemetery, and the residual probability to the current state. These are nonnegative since $h'R\le1$. Use a fixed $q'$-bit draw for the initial pair and every step, where $q'\ge1$ is the smallest integer satisfying
\begin{equation}
2^{q'}\ge\frac{1024N[N+L'(2m+2)]}{\varepsilon}.\label{eq:weightedbits}
\end{equation}
After at most $L'$ updates return the rational reward $\widehat F$.

Uniformization at rate $R$ gives Euler error $R^2/L'\le\varepsilon/(1024N)$ at horizon $1$. Lemma~\ref{lem:categorical} gives the same bound for fixed-bit sampling. The horizon is exact. Therefore, for $\nu'=\E\widehat F$,
\begin{equation}
|\nu'-\mu'|\le\varepsilon/512,\qquad0\le\widehat F\le N,\qquad
\Var(\widehat F)\le N\nu'\le2N\mu'.\label{eq:weightedfinite}
\end{equation}
Take $S'=\lceil512N/\varepsilon^2\rceil$ samples per batch and the same odd number $r$ of independent batches with $2^{-r}\le\delta$. Chebyshev's inequality at threshold $\varepsilon\mu'/4$ gives per-batch failure probability at most
\[
\frac{16\cdot2N\mu'}{S'\varepsilon^2(\mu')^2}\le\frac1{16}.
\]
Thus their median $Z_*$ satisfies, with probability at least $1-\delta$,
\begin{equation}
|Z_*-\mu'|\le\left(\frac14+\frac1{512}\right)\varepsilon\mu'
<\frac{\varepsilon\mu'}3.\label{eq:weightedmedian}
\end{equation}
The sample counts and random-bit budgets are fixed in advance; all rewards, averages, and comparisons are rational.

\subsection{A rational relative normalization}
To recover $u'$, we need relative accuracy for $e^{-\Lambda}$ even when it is exponentially small.
\begin{lemma}\label{lem:exp}
For rational $\Lambda\in[0,R]$, a positive rational $q_0$ satisfying
\begin{equation}\left|\frac{q_0}{e^{-\Lambda}}-1\right|\le\varepsilon/64\label{eq:exp}\end{equation}
can be computed in time polynomial in the bit encodings, the numerical bound $R+1$, and $\log(1/\varepsilon)$.
\end{lemma}
\begin{proof}
Use $q_0=1$ when $\Lambda=0$. Otherwise put $k=\max\{1,\lceil\Lambda\rceil\}$ and $x=\Lambda/k\in(0,1]$. We first approximate $e^{-x}$, which is bounded below by a positive constant, and then take the $k$th power. An error of order $\varepsilon/k$ in each factor will give an error of order $\varepsilon$ in the product. For
\[
a_t=\sum_{j=0}^t\frac{(-x)^j}{j!},
\]
the alternating-series remainder is at most $1/(t+1)!$. Choose the smallest $t\ge1$ for which this bound is at most $\varepsilon/(1024k)$ and set $q_0=a_t^k$. The choice has $t=O(1+\log(k/\varepsilon))$, because factorial growth eventually exceeds geometric growth. As $e^{-x}\ge e^{-1}>1/3$,
\[
a_t=e^{-x}(1+\theta),\qquad|\theta|\le\frac{3\varepsilon}{1024k}=:\alpha<1.
\]
In particular $a_t>0$. Bernoulli's inequality and $1+z\le e^z$ give
\[
1-k\alpha\le(1-\alpha)^k\le\frac{q_0}{e^{-\Lambda}}
\le(1+\alpha)^k\le e^{k\alpha}.
\]
The lower error is at most $3\varepsilon/1024<\varepsilon/64$. Since $k\alpha<1/2$, the estimate $e^z\le1/(1-z)\le1+2z$ for $0\le z\le1/2$ bounds the upper error by $6\varepsilon/1024<\varepsilon/64$. All computations are exact rational operations. Finally, $k\le R+1$ is polynomially bounded in the input length by~\eqref{eq:ratesum}, so exponentiation produces a rational number with polynomially many bits.
\end{proof}

\begin{proof}[Proof of Theorem~\ref{thm:main}]
Return the boundary answers from preprocessing, or otherwise return $\widehat u=\min\{1,q_0Z_*\}$. The three approximations enter through the factorization
\[
\frac{q_0Z_*}{u(\mathbf p)}
=\frac{Z_*}{\mu'}\cdot\frac{q_0}{e^{-\Lambda}}\cdot\frac{u'}{u(\mathbf p)}.
\]
The factors account respectively for sampling the normalized mean, approximating the exponential normalization, and rounding the failure rates. On the event~\eqref{eq:weightedmedian}, which has probability at least $1-\delta$, the ratio of the unclipped output to $u(\mathbf p)$ lies between
\[
(1-\varepsilon/3)(1-\varepsilon/64)e^{-\varepsilon/64}
\quad\text{and}\quad
(1+\varepsilon/3)(1+\varepsilon/64)e^{\varepsilon/64}.
\]
The lower bound is at least $1-(1/3+1/64+1/64)\varepsilon>1-\varepsilon$, by $e^{-a}\ge1-a$ and the product inequality for factors in $[0,1]$. The logarithm of the upper bound is at most $35\varepsilon/96$. Convexity gives
\[
e^{35\varepsilon/96}\le1+(e^{35/96}-1)\varepsilon<1+\varepsilon,
\]
where $35/96<1/2$ and $e^{1/2}<2$. Clipping at $1$ preserves the guarantee.

Preprocessing uses the explicit incidence list. Rates have polynomial encoding length, and Lemma~\ref{lem:weightedcut} computes the minimum cut exactly with polynomial bit complexity. The bound on $R$ in~\eqref{eq:ratesum} makes $L'$ polynomial in the input length and $\varepsilon^{-1}$. The bit width $q'$, batch size $S'$, and number of batches are polynomial in the required parameters. State updates use the scans of Lemma~\ref{lem:complexity}, Lemma~\ref{lem:exp} supplies rational normalization, and exact averaging has the same denominator bound as before. Thus the running time and random-bit usage are polynomially bounded for every random outcome.
\end{proof}

%% file: witnesses.tex
\section{A general formulation with surviving witnesses}
The estimator combines a path-weighting identity with a hazard bound. The following theorem states the identity in terms of surviving witnesses. In the hypergraph application, paired heads provide the hazard bound and an efficient proposal process.

\begin{theorem}\label{thm:witness}
Let $P$ be a finite-state continuous-time Markov chain with bounded rates, observed until time $T$. Let $\mathcal W$ be a finite set of $M\ge1$ witnesses, all initially alive. A witness may die at a jump and never revives; the state records the surviving set $\mathcal W(x)$. Let $\mathcal E=\{\mathcal W(X_T)\ne\varnothing\}$.

For each $w\in\mathcal W$, suppress exactly the transitions destroying $w$, retain all other rates, and denote the resulting chain by $Q_w$. Let $k_w(x)$ be the total suppressed rate. Choose $w$ uniformly, sample $Q_w$, and put
\begin{equation}
W=\frac{M}{|\mathcal W(X_T)|}\exp\!\left(-\int_0^Tk_w(X_t)\,dt\right).
\label{eq:witness}
\end{equation}
Then $\E W=\PP_P(\mathcal E)$. If every suppressed path satisfies $\int_0^T k_w(X_t)\,dt\ge A_0$, then
\[
0\le W\le Me^{-A_0},\qquad
\frac{\E W^2}{\PP_P(\mathcal E)^2}
\le\frac{Me^{-A_0}}{\PP_P(\mathcal E)}
\]
whenever $\PP_P(\mathcal E)>0$.
\end{theorem}
\begin{proof}
On a path preserving $w$, the original holding rate exceeds the suppressed holding rate by $k_w(x)$, and the factors corresponding to actual jump rates agree. The density ratio is the exponential in~\eqref{eq:witness}, including the final holding interval. Thus
\[
\E_{Q_w}\!\left[e^{-\int_0^Tk_w(X_t)\,dt}F\right]
=\E_P[\mathbf1_{\{w\in\mathcal W(X_T)\}}F].
\]
A finite-state chain with bounded rates almost surely makes only finitely many jumps on a finite interval. Summing over jump sequences is therefore valid: use monotone convergence for nonnegative functionals and subtraction for bounded signed ones. Take $F=1/|\mathcal W(X_T)|$ on paths retaining $w$ and average over $w$. The pathwise identity
\[
\sum_{w\in\mathcal W}\frac{\mathbf1_{\{w\in\mathcal W(X_T)\}}}{|\mathcal W(X_T)|}
=\mathbf1_{\mathcal E},
\]
with zero on paths with no survivors, proves unbiasedness. The assumed lower bound on the integrated hazard yields the range bound, and $W^2\le Me^{-A_0}W$ gives the second moment.
\end{proof}

For hypergraphs, witnesses are original representative pairs, so $M=N$ and $|\mathcal W(X_T)|=\binom{K(X_T)}2$ when $K\ge2$. Paired heads supply $A_0=\lambda T$. The minimum-cut bound $u_H(p)\ge e^{-A_0}$ then gives polynomial relative variance. The proposal, terminal multiplicity, and normalization are all computable from a polynomial-size state. Together, these properties turn the identity into an approximation algorithm.

\section{Further directions}\label{sec:discussion}
The construction establishes polynomial-time relative approximation for all-terminal disconnection in explicit hypergraphs with independent rational edge failures. Its sampling law depends on four structural properties: a head rule common to all protected pairs, persistent availability of suppressed copies, unchanged safe rates, and likelihood accumulation through the terminal horizon. The final number of representatives determines the multiplicity correction.

The rational Euler implementation gives an explicit worst-case bound on bit complexity. A faster implementation could reduce the discretization cost or use an event-driven sampler with a fixed precision budget. Other connectivity events would require witness families and hazard certificates suited to their own structure. The general identity in Theorem~\ref{thm:witness} makes these requirements explicit.

Appendix~\ref{app:checks} gives an exact recurrence that verifies the likelihood and multiplicity identities on finite instances.

\appendix
\section{Exact polynomial checks on small instances}\label{app:checks}
Fix a protected pair and safe state $x$. Let $\ell(x)$ count original nonloop labels and define
\[
F_x(t)=\E_{P,x}\!\left[\frac{\mathbf1_{\{w\text{ survives to }t\}}}{\binom{K(X_t)}2}\right],
\]
with the integrand zero whenever the pair does not survive, including paths ending with one block. Its backward equation is
\begin{equation}
F_x'(t)=-\ell(x)F_x(t)+
\sum_{x\xrightarrow{\text{safe }c}y}\frac12F_y(t),
\qquad F_x(0)=\frac1{\binom{K(x)}2}.
\label{eq:recurrence}
\end{equation}
The sum retains safe copy labels. Every effective contraction turns its contracted original edge into a loop, so $\ell(y)<\ell(x)$ for every successor in the sum. Inductively, $F_x(t)$ is a rational linear combination of $e^{-jt}$ for integers $0\le j\le\ell(x)$.

More explicitly, suppose $F_y(t)=\sum_j a_{y,j}e^{-jt}$ is known for each safe successor. Assign the initial coefficient $1/\binom{K(x)}2$ to $e^{-\ell(x)t}$ and, for each safe copy and each successor coefficient, add
\begin{equation}
\frac{a_{y,j}}{2[\ell(x)-j]}
\left(e^{-jt}-e^{-\ell(x)t}\right).
\label{eq:polynomial}
\end{equation}
The denominator is positive because $j\le\ell(y)<\ell(x)$. Variation of constants proves~\eqref{eq:recurrence}. At $t=T$, replace $e^{-jT}$ by $p^j$ to obtain a polynomial in $p$ with rational coefficients.

Summing the initial-state polynomial over all unordered pairs gives one expression for unreliability. Independently, enumerate surviving subsets $A\subseteq E$ and add
\[
(1-p)^{|A|}p^{m-|A|}
\]
whenever $A$ has at least two components. Equality of the coefficient vectors verifies both the likelihood and final-pair identities for every $p$ on that hypergraph. The recurrence also applies to disconnected inputs with $\lambda=0$.

For positive integer weights $a_e$, replace $\ell(x)$ by the total nonloop weight and each copy rate $1/2$ by $a_e/2$. This checks the weighted process with $p_e=p^{a_e}$. Total nonloop weight decreases strictly at every safe contraction, so the denominators in the recurrence remain positive. An exact-arithmetic implementation evaluates both recurrences and checks the hazard at every state reached for a protected pair.

The implementation verifies all $2{,}048$ simple hypergraphs on four labeled vertices with edges of size at least two, $100$ fixed-seed small multihypergraphs with unit weights, and $100$ with weights in $\{1,2,3,4\}$. Across these families, it visits $77{,}719$ states with a protected pair. Every hazard comparison holds, and every pair-sum polynomial agrees coefficient by coefficient with direct enumeration. The computational record includes the random seed and the number of states examined.

\section*{Acknowledgments}
The author used OpenAI Codex for exposition, elaborating and checking proof details, reference verification, and LaTeX preparation, and takes full responsibility for the manuscript.